\documentclass[letter,tradiabstract,longauth]{aa} 
\usepackage{graphicx}
\usepackage{txfonts}
\usepackage{natbib}
\bibpunct{(}{)}{;}{a}{}{,} 

\def\gtrsim{\mathrel{\hbox{\rlap{\hbox{\lower4pt\hbox{$\sim$}}}\hbox{$>$}}}}
\def\lesssim{\mathrel{\hbox{\rlap{\hbox{\lower4pt\hbox{$\sim$}}}\hbox{$<$}}}}
\newcommand{\hi}{H{\sc i}}

\newcommand{\halfa}{H$\alpha$$\,$}

\newcommand{\oii}{[O{\sc ii}]$\,$}

\newcommand{\msun}{M$_{\odot}$$\,$}

\newcommand{\arsec}{$^{\prime\prime}\!$}
\newcommand{\D}{{$\mathcal D$} }
\newcommand{\micron}{$\mu$m}
\newcommand{\hers}{{\it Herschel}}

\begin{document}

   \title{The  {\em Herschel} Virgo cluster survey: V. Star-forming dwarf galaxies - dust in metal-poor environments\thanks{{\it Herschel} is an ESA space observatory with science instruments provided
by European-led Principal Investigator consortia and with important participation from NASA.}}

\author{
M. Grossi\inst{1}
\and
L. K. Hunt\inst{2}
\and
S. Madden\inst{3}
\and
C. Vlahakis\inst{4}
\and
D. J. Bomans\inst{5}
\and
M. Baes\inst{6}
\and
G. J. Bendo\inst{7}
\and
S. Bianchi\inst{2}
\and
A. Boselli\inst{8}
\and
M. Clemens\inst{9}
\and
E. Corbelli\inst{2}
\and
L. Cortese\inst{10}
\and
A. Dariush\inst{10}
\and
J. I. Davies\inst{10}
\and
I. De Looze\inst{6}
\and
S. di Serego Alighieri\inst{2}
\and
D. Fadda\inst{11}
\and
J. Fritz \inst{6}
\and
D. A. Garcia-Appadoo\inst{12}
\and
G. Gavazzi\inst{13}
\and
C. Giovanardi\inst{2}
\and
T. M. Hughes\inst{10}
\and
A. P. Jones\inst{14}
\and
D. Pierini\inst{15}
\and
M. Pohlen\inst{10}
\and
S. Sabatini\inst{16}
\and
M. W. L. Smith\inst{10}
\and
J. Verstappen\inst{6}
\and
E. M. Xilouris\inst{17}
\and
S. Zibetti\inst{18}
}

\institute{
CAAUL, Observat\'orio Astron\'omico de Lisboa, Universidade de Lisboa, Tapada da Ajuda, 1349-018, Lisboa, Portugal
\email{grossi@oal.ul.pt}
\and
INAF-Osservatorio Astrofisico di Arcetri, Largo Enrico Fermi 5, 50125 Firenze, Italy
\and
Laboratoire AIM, CEA/DSM- CNRS - Universit\'e Paris Diderot, Irfu/Service d'Astrophysique, 91191 Gif sur Yvette, France
\and
Leiden Observatory, Leiden University, P.O. Box 9513, NL-2300 RA Leiden, The Netherlands
\and
Astronomical Institute, Ruhr-University Bochum, Universitaetsstr. 150, 44780 Bochum, Germany
\and
Sterrenkundig Observatorium, Universiteit Gent, Krijgslaan 281 S9, B-9000 Gent, Belgium
\and
Astrophysics Group, Imperial College London, Blackett Laboratory, Prince Consort Road, London SW7 2AZ, UK
\and
Laboratoire d'Astrophysique de Marseille, UMR 6110 CNRS, 38 rue F. Joliot-Curie, F-13388 Marseille, France
\and
INAF-Osservatorio Astronomico di Padova, Vicolo dell'Osservatorio 5, 35122 Padova, Italy
\and
Department of Physics and Astronomy, Cardiff University, The Parade, Cardiff, CF24 3AA, UK
\and
NASA Herschel Science Center, California Institute of Technology, MS 100-22, Pasadena, CA 91125, USA
\and
ESO, Alonso de Cordova 3107, Vitacura, Santiago, Chile
\and
Universita' di Milano-Bicocca, piazza della Scienza 3, 20100, Milano, Italy
\and
Institut d'Astrophysique Spatiale (IAS), Batiment 121, Universite Paris-Sud 11 and CNRS, F-91405 Orsay, France
\and
Max-Planck-Institut fuer extraterrestrische Physik, Giessenbachstrasse, Postfach 1312, D-85741, Garching, Germany
\and
INAF-Istituto di Astrofisica Spaziale e Fisica Cosmica, via Fosso del Cavaliere 100, I-00133, Roma, Italy
\and
Institute of Astronomy and Astrophysics, National Observatory of Athens, I. Metaxa and Vas. Pavlou, P. Penteli, GR-15236 Athens, Greece
\and
Max-Planck-Institut fuer Astronomie, Koenigstuhl 17, D-69117 Heidelberg,  Germany
}



\abstract{We present the dust properties of a small sample of Virgo cluster
dwarf galaxies drawn from the science demonstration phase data set of the
{\em Herschel} Virgo Cluster Survey. These galaxies have low metallicities (7.8 $<$ 12 + log(O/H) $<$ 8.3) and star-formation rates $\lesssim$ 10$^{-1}$ \msun yr$^{-1}$. We measure the spectral energy distribution (SED) from 100 to 500 \micron\ and derive dust temperatures and dust masses. The SEDs are fitted by a cool component of temperature T $\lesssim$ 20 K, implying dust masses around 10$^{5}$ \msun and dust-to-gas ratios \D  within the range 10$^{-3}$-10$^{-2}$.
The completion of the full survey will yield a larger set of galaxies, which will provide more stringent constraints on the dust content of star-forming dwarf galaxies.}


\maketitle

%

\section{Introduction}

Late-type dwarf galaxies are metal-poor systems, and they represent unique environments to investigate the properties of dust in the low-metallicity regime.
Mid/far-infrared (MIR/FIR) and submillimetre (submm) observations have provided information about the different dust
components in dwarfs, showing that even metal-poor galaxies may host a significant amount of dust \citep{1999ApJ...516..783T,2004ApJS..154..211H,2008ApJ...678..804E}.
While the majority of the studies to date have focused mainly on bright and isolated dwarfs \citep{2003A&A...407..159G,2005A&A...434..849H,2005A&A...434..867G,2009A&A...508..645G},
little is known about the interplay between the environment and the dust properties of low-mass and low-metallicity systems.

It is still unclear whether most of the dust mass in these systems is at very low temperature (T $\lesssim$ 10 K), as the excess emission in the submm observed in some dwarfs might suggest \citep{2002Ap&SS.281..247M,2005A&A...434..867G,2009A&A...508..645G}. This cold dust component may either be associated with the star-forming regions, residing in clumpy molecular complexes \citep{2003A&A...407..159G}, or extend beyond the optical disc following the distribution of the neutral hydrogen \citep[\hi;][]{2002ApJ...567..221P}.
However, the main objection to this interpretation is the high dust mass associated with the cold dust, and alternative explanations have been proposed
\citep[see][and references therein]{2010MNRAS.402.1409B}. For example, \citet{2002A&A...382..860L}  used a mixed dust model \citep{1990A&A...237..215D} of large and small grains, finding that a high abundance of small grains (compared to the Milky Way) with a shallow emissivity ($\beta < 2$) may account for the excess submm emission in NGC\,1569. \citet{2009ApJ...706..941Z} applied a similar approach to  the starburst spiral galaxy NGC\,3310  without introducing a cold dust component.
Only additional measurements in the FIR/submm regime and the analysis of a larger sample of galaxies can help us properly assess the issue of dust at low temperatures.

The \hers\, Space Observatory \citep{Pilbratt2010} allows us for the first time to  study the wavelength range between 200  and 500 \micron, where the emission from cold dust is expected to be the most prominent.
As part of the \hers\, Open Time Key Project, the \hers\, Virgo Cluster Survey\footnote{http://www.hevics.org} \citep[HeViCS;][]{Davies2010} will map an area of 64 square degrees of the Virgo cluster with PACS \citep{Poglitsch2010} and SPIRE \citep{Griffin2010} to investigate the dust content of the different morphological types within the cluster.
The dwarf galaxy population of the Virgo cluster is dominated by dwarf ellipticals, but also contains a non-negligible fraction ($\sim$ 10\%) of late-type dwarfs with signs of current star formation activity \citep{1987AJ.....94..251B}.
In this Letter, we present the dust properties of three of these galaxies detected with PACS and SPIRE in the HeViCS Science Demonstration Phase (SDP) data set, a 4$^{\circ} \times 4^{\circ}$ field covering the central region of the cluster.

   \begin{figure*}
   \centering
   \includegraphics[bb=14 14 300 300,width=3.cm,clip]{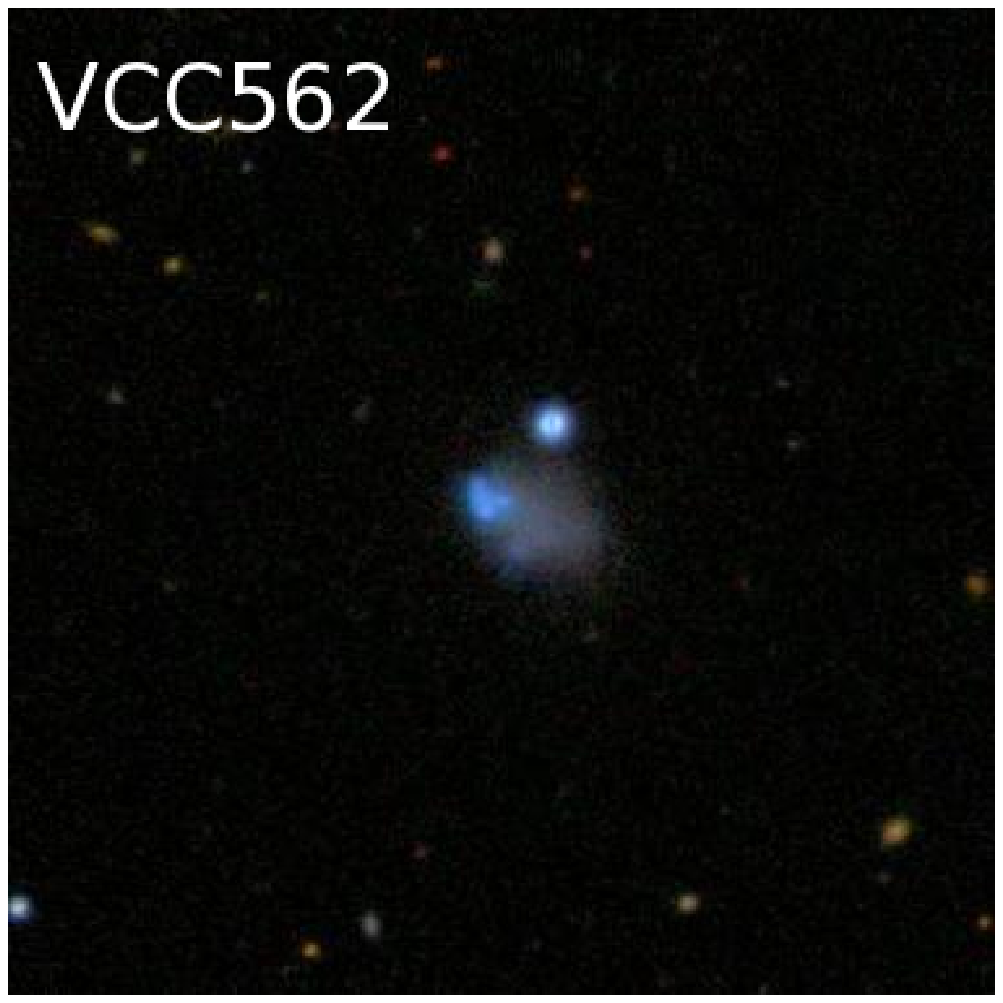}
    \includegraphics[bb=54 360 270 576,width=3cm,clip]{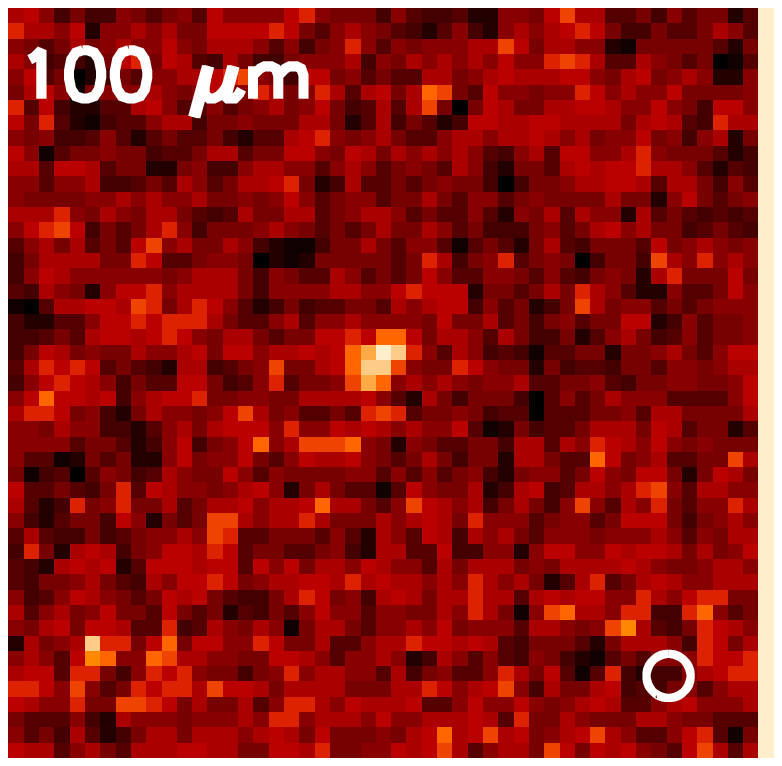}
    \includegraphics[bb=54 360 270 576,width=3cm,clip]{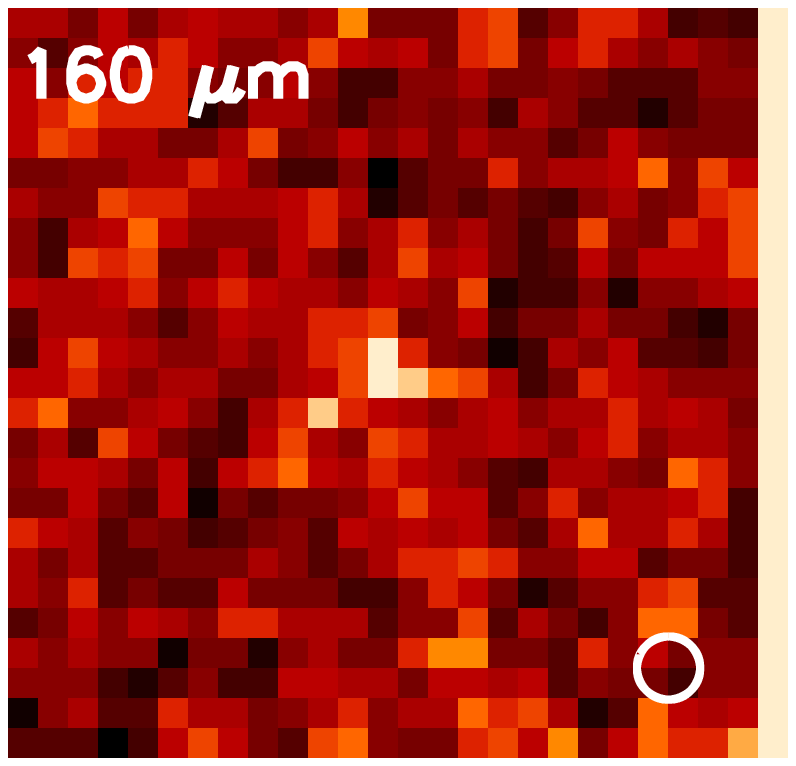}
   \includegraphics[bb=54 360 270 576,width=3cm,clip]{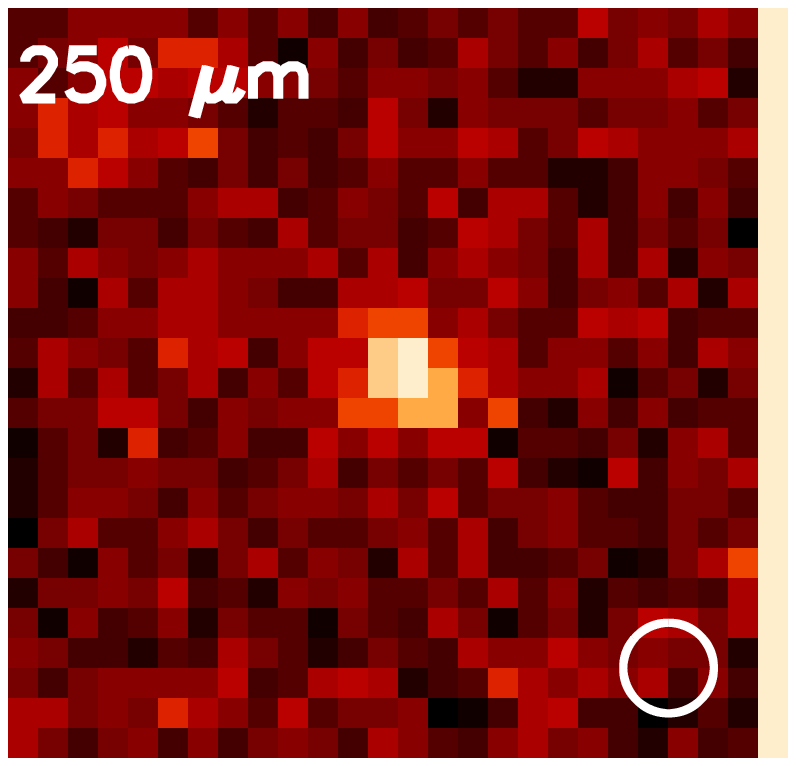}
   \includegraphics[bb=54 360 270 576,width=3cm,clip]{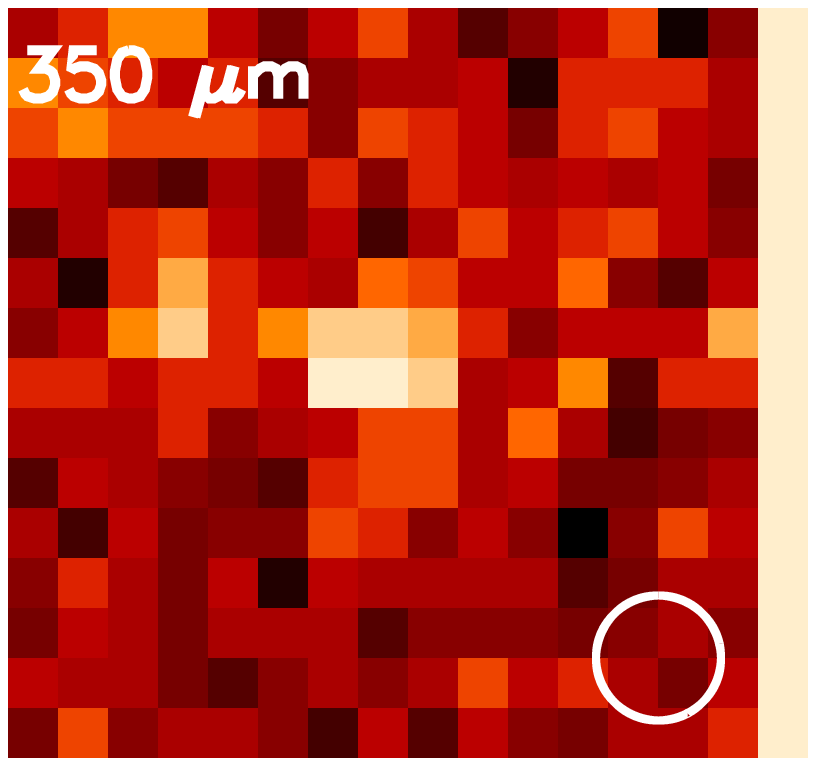}
   \includegraphics[bb=54 360 270 576,width=3cm,clip]{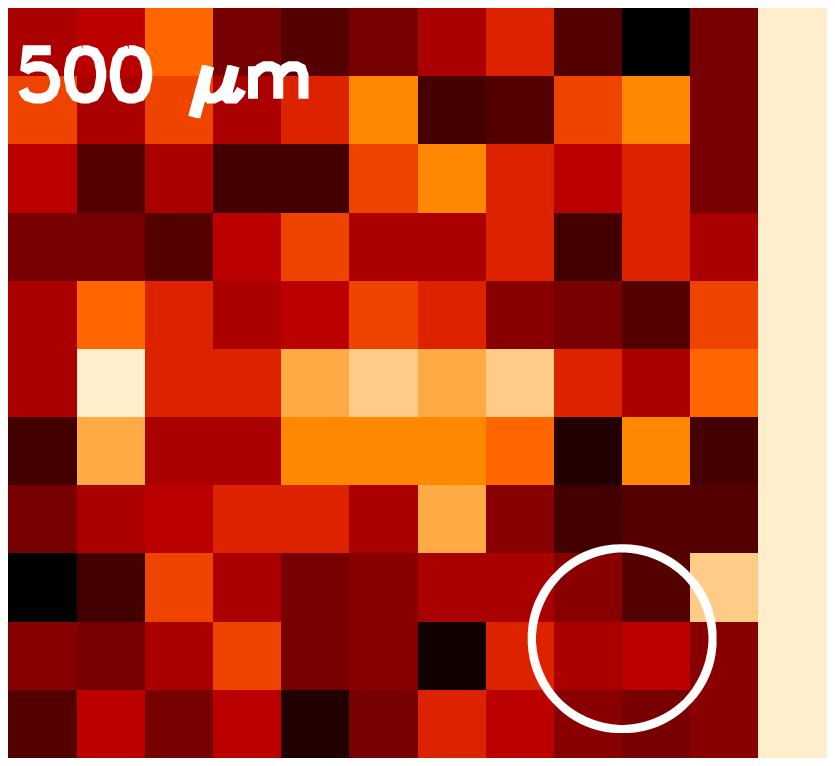} \\
   \includegraphics[bb=14 14 300 300,width=3.cm,clip]{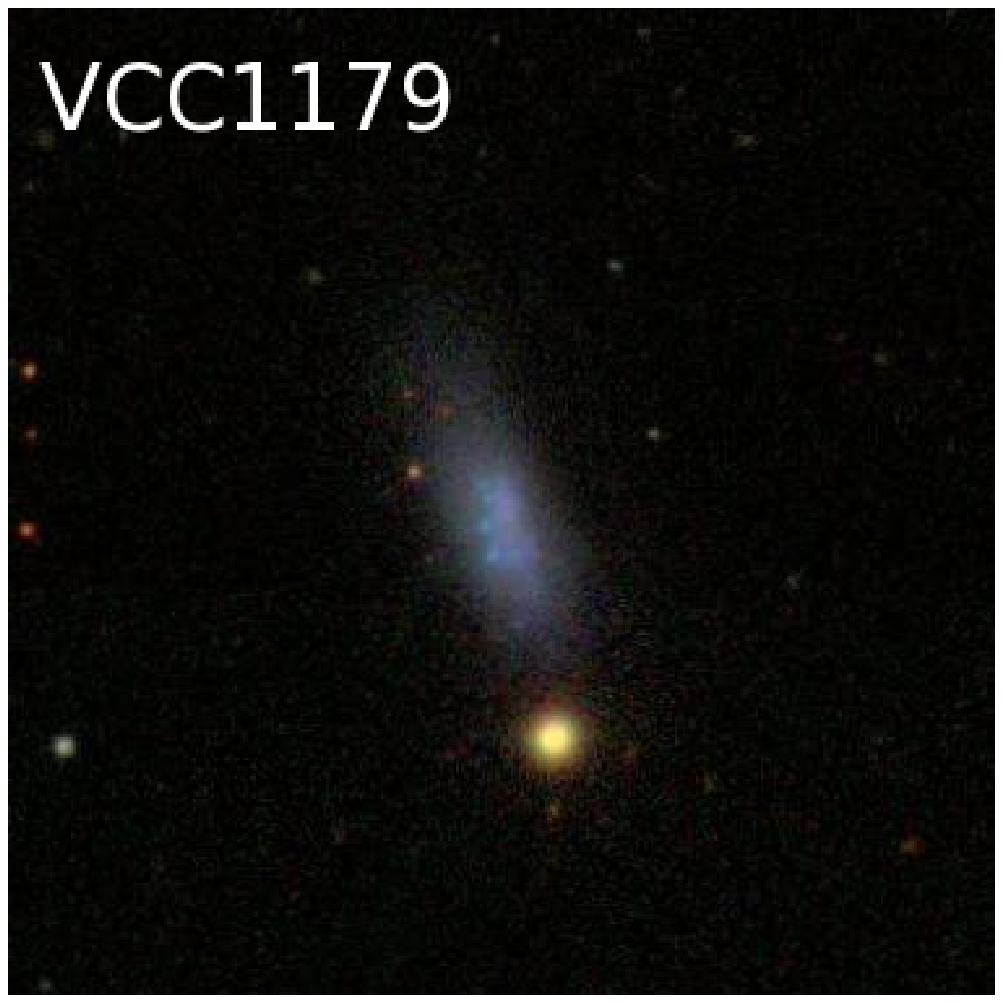}
   \includegraphics[bb=54 360 270 576,width=3cm,clip]{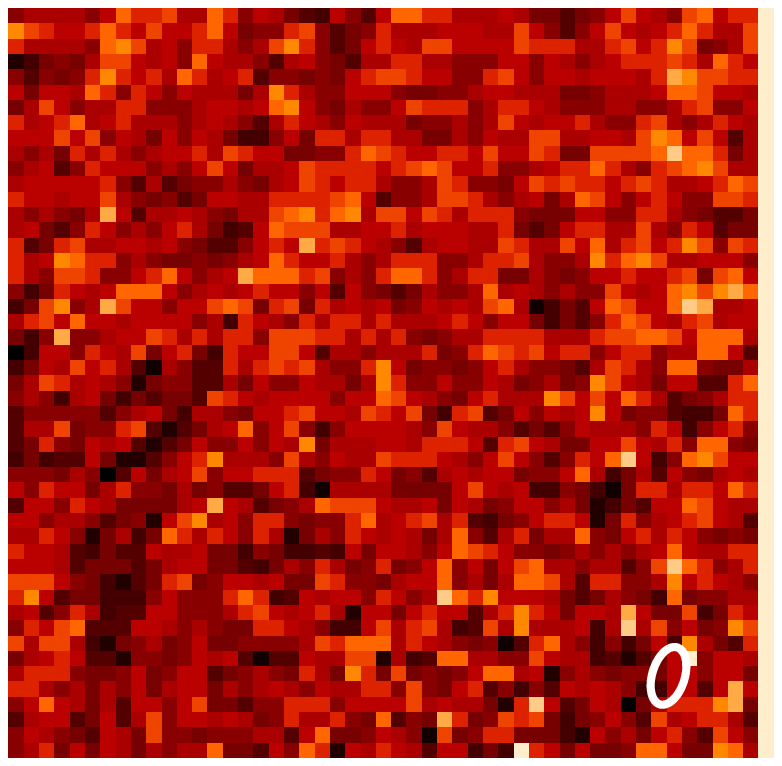}
   \includegraphics[bb=54 360 270 576,width=3cm,clip]{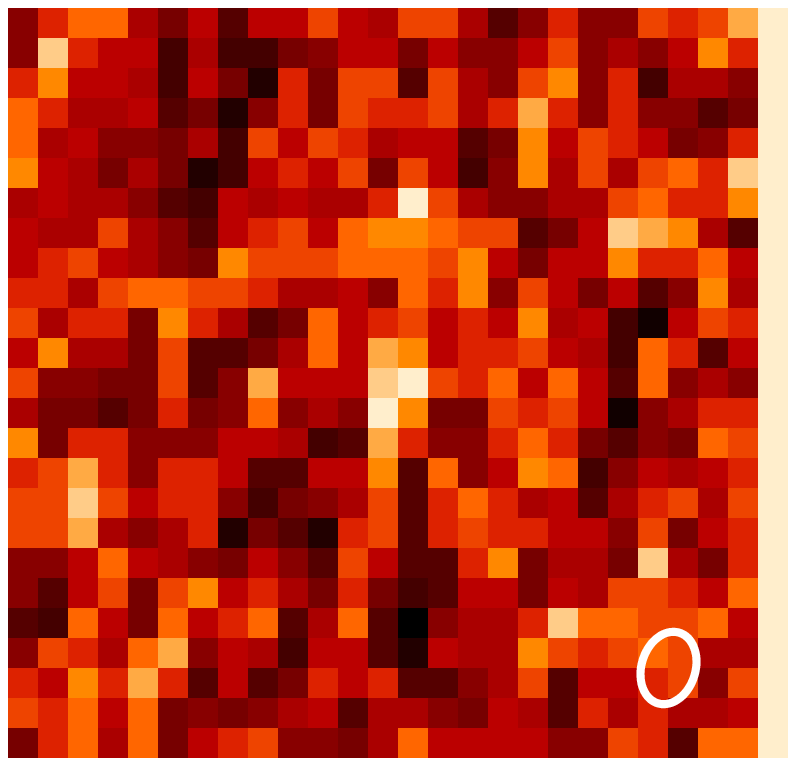}
   \includegraphics[bb=54 360 270 576,width=3cm,clip]{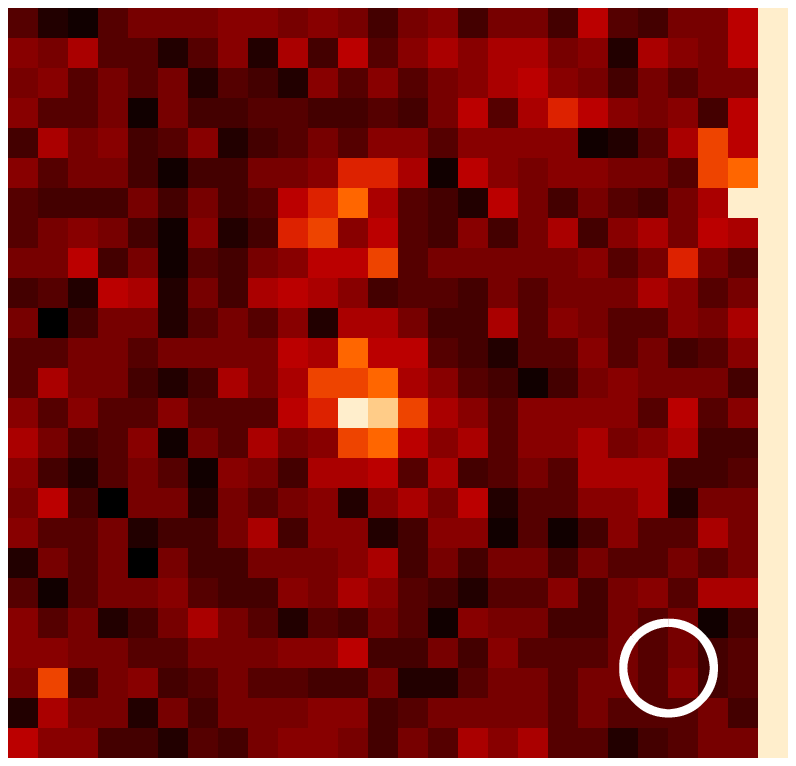}
   \includegraphics[bb=54 360 270 576,width=3cm,clip]{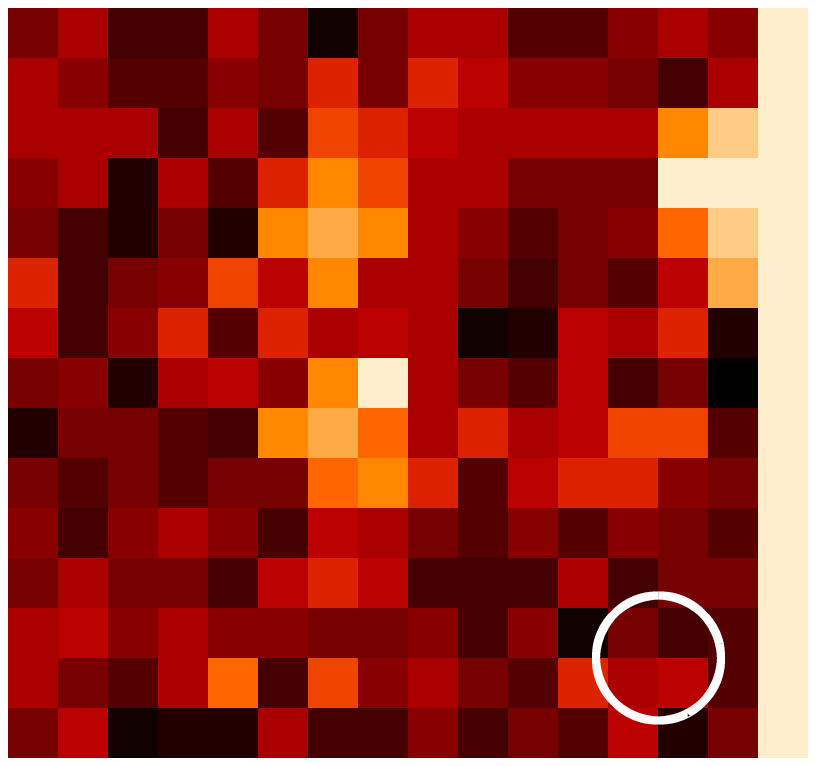}
   \includegraphics[bb=54 360 270 576,width=3cm,clip]{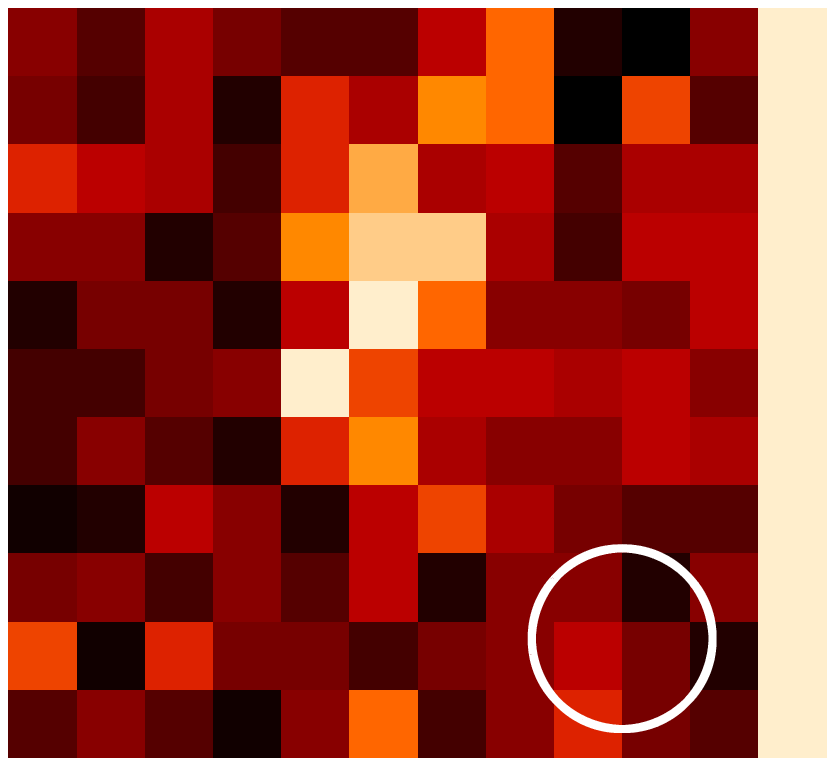} \\
   \includegraphics[bb=14 14 300 300,width=3.cm,clip]{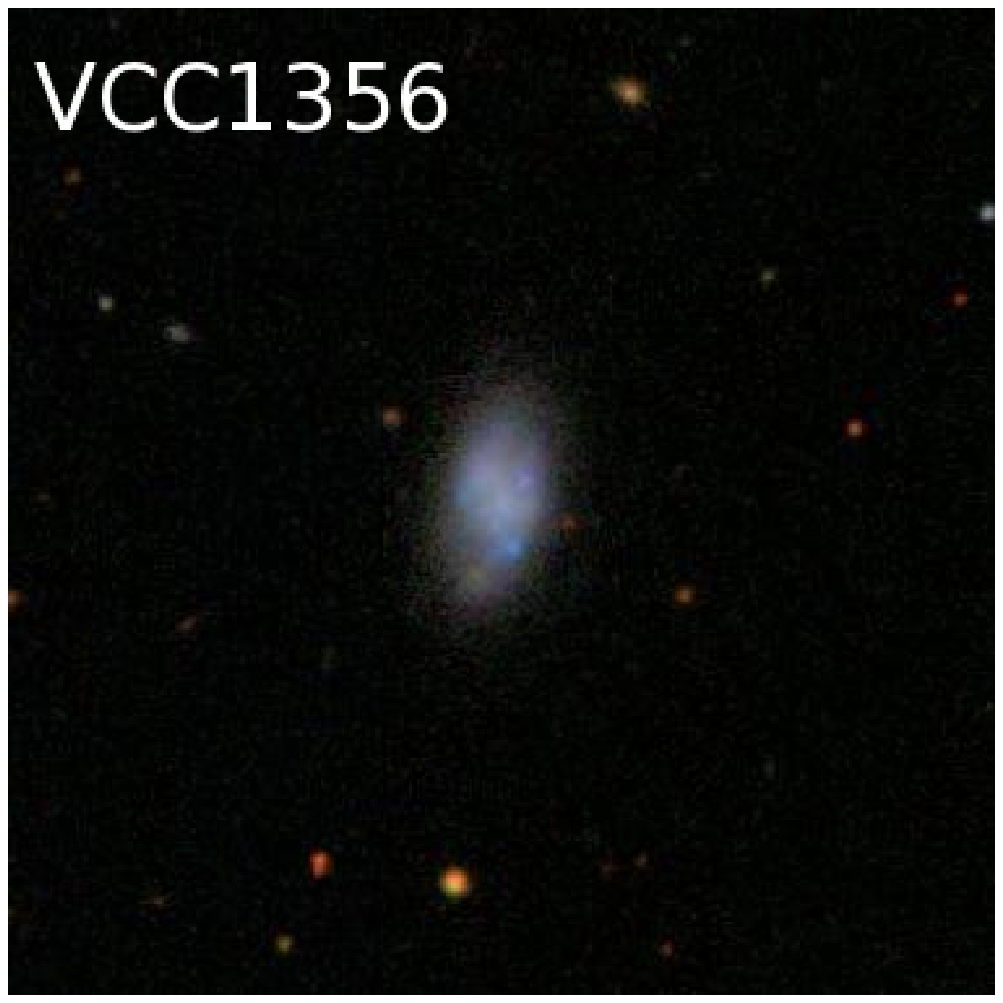}
   \includegraphics[bb=54 360 270 576,width=3cm,clip]{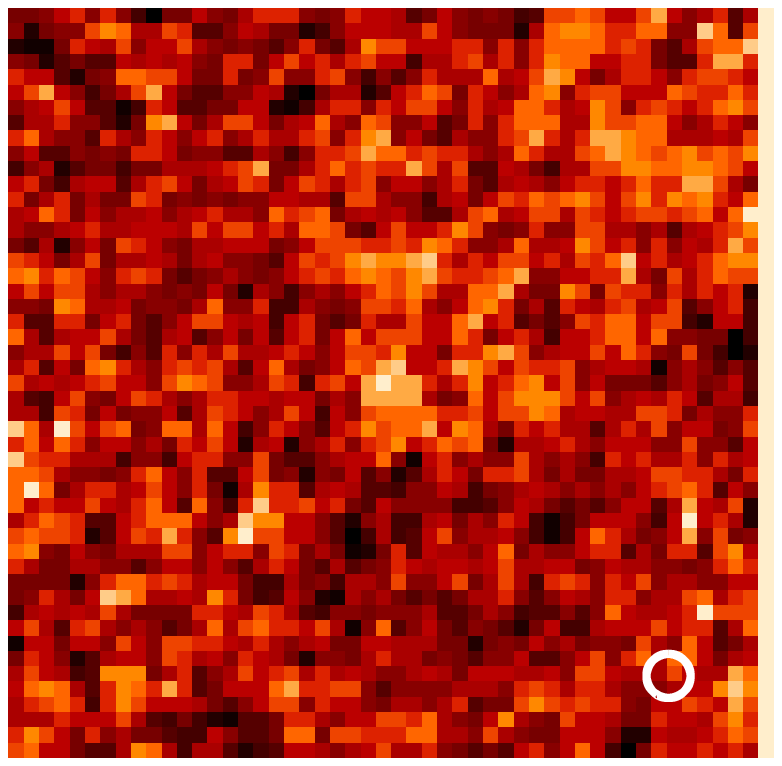}
   \includegraphics[bb=54 360 270 576,width=3cm,clip]{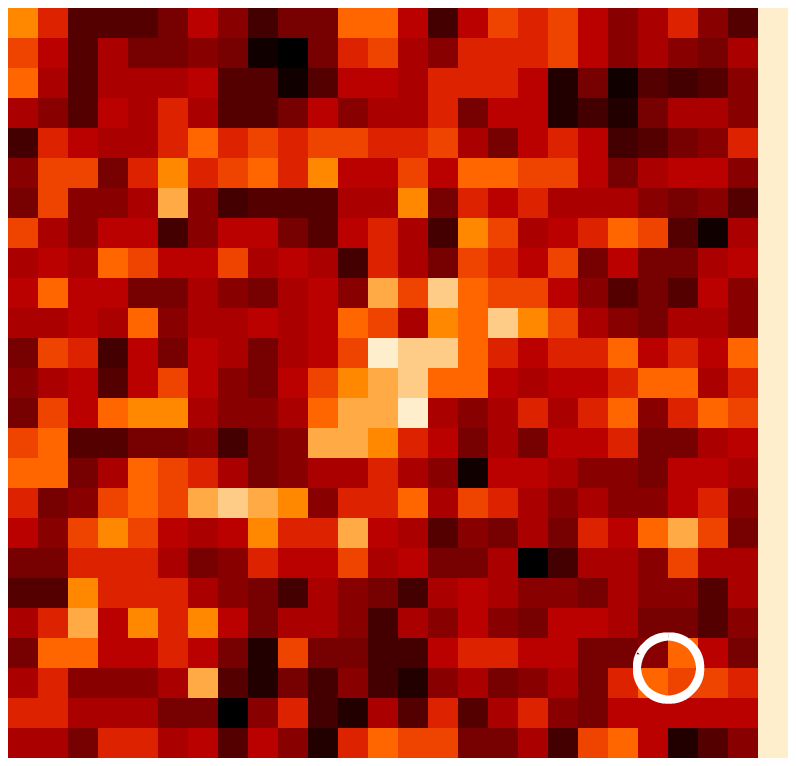}
   \includegraphics[bb=54 360 270 576,width=3cm,clip]{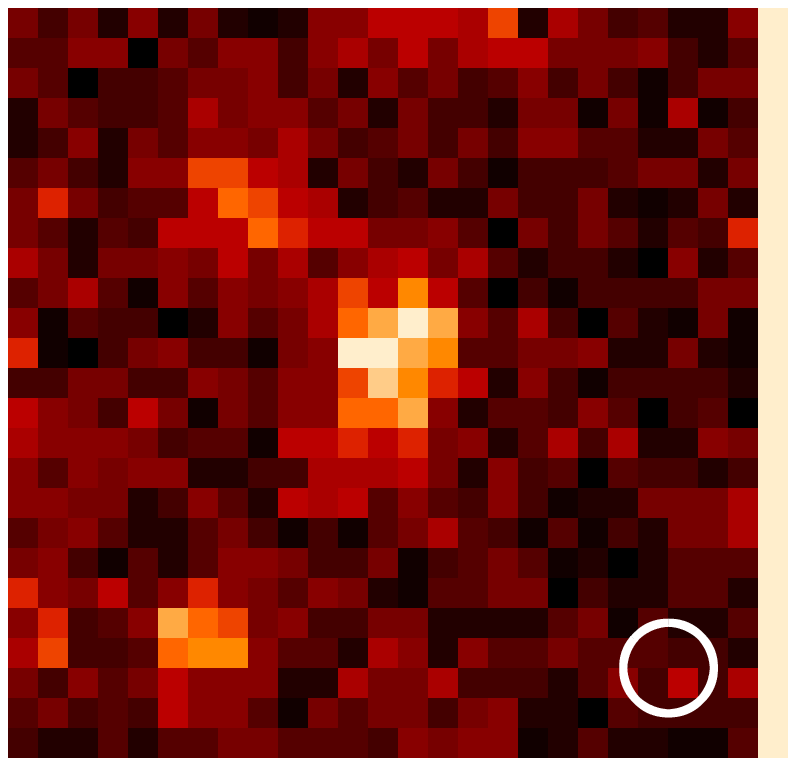}
   \includegraphics[bb=54 360 270 576,width=3cm,clip]{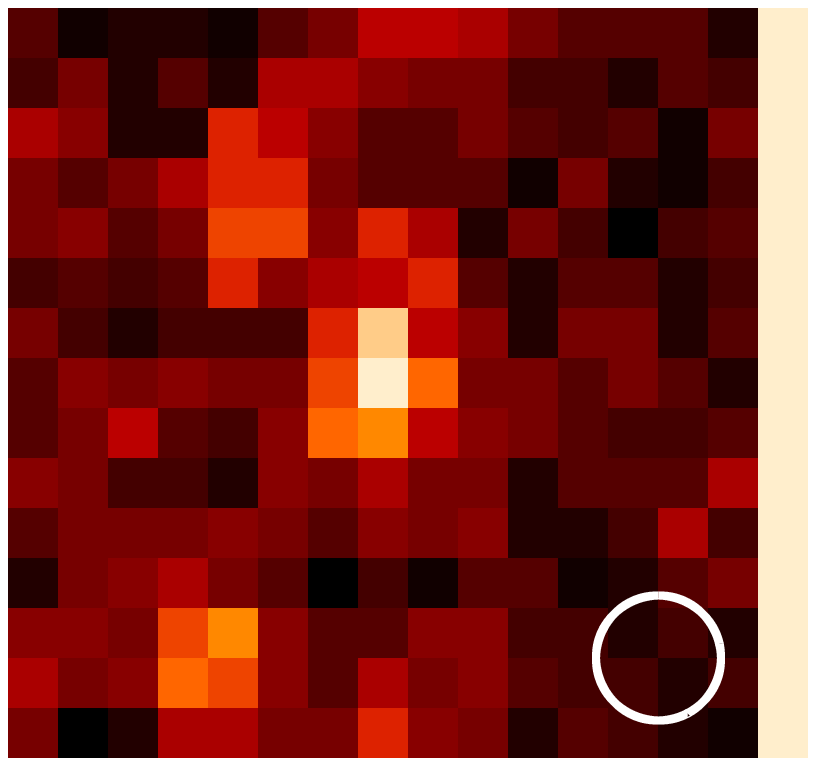}
   \includegraphics[bb=54 360 270 576,width=3cm,clip]{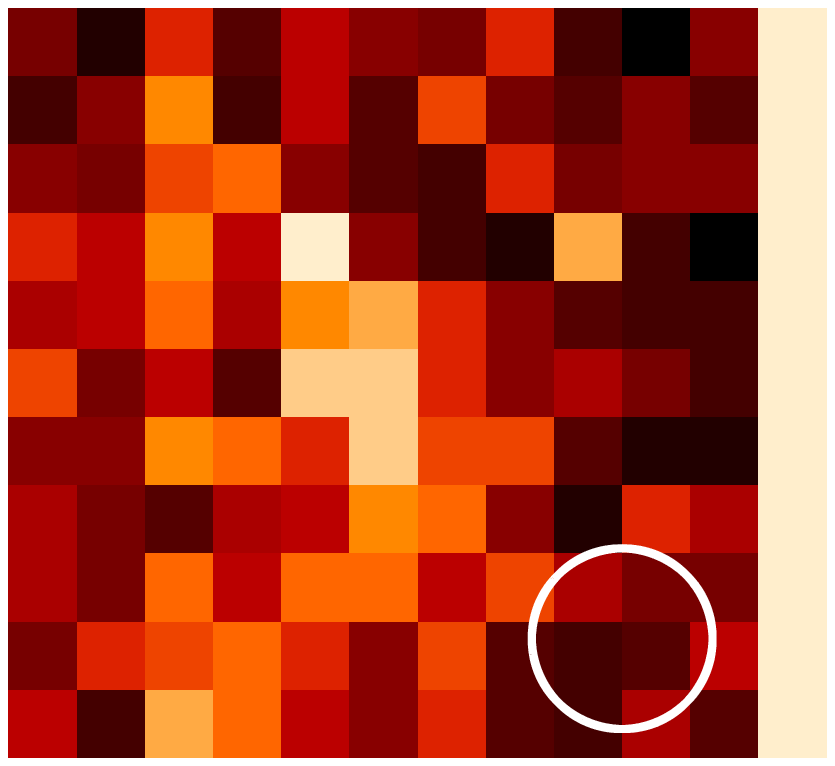}
   \caption{SDSS and \hers \, images of VCC\,562, VCC\,1179, and VCC\,1356 at 100, 160, 250, 350, and 500 \micron. The field size is 150\arsec. The FWHM of the beam at each wavelength is displayed at the bottom-right corner. The different shape of the PACS beams depend on the single-scan (ellipse) or crossed-link scan (circle) coverage of the area.}
              \label{fig:sdssspire}%
    \end{figure*}

\section{Sample selection}

The HeViCS SDP field contains 28 late-type low-luminosity
galaxies classified in the Virgo Cluster Catalog \citep[VCC,][]{1985AJ.....90.1681B} as Sm (7), Im (12), blue compact dwarfs (BCDs) (6), and dwarf irregulars (3). Three of the BCDs  have been detected with PACS and
SPIRE and are analysed in this work: VCC\,562, VCC\,1179, and VCC\,1356. They have
$B$ magnitudes between 15.5 and 16.7
($M_B$ between $-14.5$ and $-15.6$),
and \hi\ masses of 10$^7$ $-$ 10$^8$ \msun assuming a distance to Virgo of 16.5 Mpc \citep{2007ApJ...655..144M}.
We inferred star-formation rates (SFRs) of between 10$^{-1}$ and 10$^{-2}$ \msun yr$^{-1}$ from extinction-corrected \halfa and UV fluxes  \citep{2004A&A...417..499G, 2009ApJ...706.1527B}.
Although these dwarfs are compact, thus classified as BCDs by
\citet{1985AJ.....90.1681B}, their SFRs are lower than the typical values for
BCDs \citep{2004AJ....128.2170H,2009MNRAS.399..924Z}.
Metallicities were estimated using the Sloan Digitised Sky Survey (SDSS)
spectroscopic database \citep{2000AJ....120.1579Y}.  Since the \oii$\lambda$3727 nebular line is outside the observed
wavelength range of the SDSS spectra, we estimated the oxygen abundances
following the method of \citet{2007ApJ...669..299P}. 
Our estimates agree, within the errors, with the measurements available in the literature for VCC\,562\footnote{Note that according to \citet{2003ApJS..145..225V} the uncertainties in the measurement of VCC\,562 are also compatible with a higher metal abundance, 12 + log(O/H)=8.4-8.6.} and VCC\,1179 \citep{2003ApJS..145..225V}.
The basic parameters of the three galaxies are given in Table 1. 

Among the Sm and Im types, only one bright IBm galaxy, IC3583 (VCC\,1686), was detected in
the SDP field, but given its different properties from the three
dwarfs, it will be analysed in more detail in a separate work.

\section{Photometry}


\begin{table*}
\begin{minipage}{\textwidth}
\label{tab:data}
\caption{The properties of the three star-forming dwarf galaxies in the Virgo cluster under study. Dust temperature, masses, and dust-to-gas ratios were obtained by fitting a single modified black-body function with $\beta =$ 2. The distance to Virgo is assumed to be 16.5 Mpc.}
\begin{center}
\begin{tabular}{lccccccc}
\hline \hline
ID         &   m$_B$    &12+log(O/H) &  SFR  &    log(M$_{HI}$/M$_{\odot}$) & T$_{d}^{1BB}$ & log(M$_{d}$/M$_{\odot}$) &  log(\D)\\
           &    mag     &            &\msun yr$^{-1}$ &    &  K         &    &              \\
\hline \hline
\object{VCC\,562}     &  16.74     &    7.8$\pm0.2$    &   4.6 $\times 10^{-2}$ & 7.67$\pm$0.03  & 20.2$\pm$0.2 &  5.18$\pm$0.13 & -2.49$\pm$0.13 \\
\object{VCC\,1179}    &  15.46     &    8.3$\pm0.2$    &   9.4 $\times 10^{-2}$ & 7.39$\pm0.06$  & 16.2$\pm0.6$ &  5.48$\pm$0.14 & -1.91$\pm$0.15  \\
\object{VCC\,1356}    &  16.15     &    8.0$\pm0.3$    &   4.6 $\times 10^{-2}$ & 8.35$\pm$0.01  & 17.9$\pm1.1$ &  5.53$\pm$0.16 & -2.82$\pm$0.16 \\
\hline \hline
\end{tabular}
\end{center}
\end{minipage}
\end{table*}


The SDP data were reduced using the Level I procedures described in \citet{Pohlen2010}. More details on the
observations can be found in \citet{Davies2010}.
The angular resolution for PACS in fast scan parallel mode is 7\arsec$\times$12\farcs7 and 11\farcs6$\times$15\farcs7,
at 100, and 160 $\mu$m, respectively. For SPIRE, the PSF FWHM is 18\farcs1, 25\farcs2,  and
36\farcs9 at 250, 350, and 500 \micron\, respectively.
Pixel sizes are 3\farcs2, 6\farcs4, 6\arsec, 10\arsec, and 14\arsec$\,$ for the five bands.
The rms levels of empty sky regions at 100, 160, 250, 350, and 500 \micron\ are
$\sim$2\,mJy\,pix$^{-1}$, 
$\sim$5\,mJy\,pix$^{-1}$, 
$\sim$8\,mJy\,beam$^{-1}$, 
$\sim$7\,mJy\,beam$^{-1}$, and 
$\sim$9\,mJy\,beam$^{-1}$, 
respectively. 

Figure \ref{fig:sdssspire} compares PACS and SPIRE data of the three dwarfs to the  SDSS images. VCC\,1179 is not detected at 100 $\mu$m, nor VCC\,562 at 500 $\mu$m.
The other galaxies are marginally detected at 500 \micron\ with a low signal-to-noise ratio (S/N) (4$\sigma$ and 5$\sigma$ for VCC\,1179 and VCC\,1356, respectively).
The VCC\,1179 images show an additional feature to the north extending beyond the edge of the optical disc. It is not clear whether this feature is associated with the galaxy or not, but it also does not seem to be related to any background source in the SDSS images.

Photometry from 100 to 500 \micron\ was derived by means of the standard growth curve analysis. The aperture is centred on the brightness peak,
and the sky level is determined by considering the value that minimizes
the radial variation in successively larger apertures. 
At large radii, the photometry should be relatively constant,
and the asymptotic value corresponds to the total flux.
The value of the sky thus derived is always within 1.5 standard deviations of the sky
background measured in empty regions around the galaxies.
The uncertainty in the flux density is assumed to be $30$\% \citep{Boselli2010,Swinyard2010}. 


\section{Dust temperatures and masses}

To derive the temperatures of the dust, we fitted the
PACS$+$SPIRE SEDs with a single modified Planck function and an
emissivity law $k_{\nu} \propto \nu^{2}$ 
(Fig. 2, upper panels). We constrained the dust masses 
using the 250 \micron\ flux densities, and   
the emissivity $k_{\nu} = $  4.67 cm$^2$ g$^{-1}$ at 250 \micron\
\citep{2001ApJ...554..778L}.
The resulting dust masses are approximately $10^5$ \msun, and the dust temperatures are around 20 K (see Table 1).

However, Fig. 2 shows that 500 \micron\ fluxes of VCC\,1179 and VCC\,1356 tend to be underestimated
by the single-temperature fits.
This difference could be due to either thermal or non-thermal radio emission
\citep{1992ARA&A..30..575C,2005A&A...434..849H}, an additional cold ($\sim$10\,K) dust component
\citep{2003A&A...407..159G,2009A&A...508..645G}, an enhanced abundance of small grains \citep{2002A&A...382..860L}, or the different optical properties of the amorphous dust grains \citep{2007A&A...468..171M}.

We first considered the possibility of radio emission.
Following \citet{1992ARA&A..30..575C}, we estimated the non-thermal and thermal
radio flux at 500\,\micron\ that would be expected given the SFR of each galaxy.
Because of the steep frequency fall-off of the non-thermal radio component toward
higher frequencies ($\sim\nu^{-0.8}$), the estimated thermal flux at 500\,\micron\ is
higher than the non-thermal one by a factor of $\sim$9.
Nevertheless, in both cases, given the low SFR of these BCDs  ($\lesssim 10^{-1}$ \msun yr$^{-1}$), the
expected radio flux at 500\,\micron\ is of the order of 200 $\mu$Jy, 
and thus cannot be responsible for the 500 \micron\ excess.

The emission from an additional cold
dust component, $\sim$10\,K,  would peak between 200 and 300\,\micron\
and could cause the excess we observe at 500 \micron.
To explore this possibility, we refined the fiducial SED model by adding a second modified black body at a lower temperature.
For this two-component grey-body fit, we also used $\beta=2$.
Since we have four points in the SED of VCC\,1179 and the uncertainties in the flux densities are large, the fit could not be tightly constrained, and we discuss only the results obtained for VCC\,1356.
More than one set of parameters were able to provide a reasonable fit to the SED of this galaxy. We inferred mean
temperatures of T$_c$ = 8\,K and T$_w$ = 19.1\,K for the two dust components and the fit for these values is displayed in Fig. 2 (lower panel). The corresponding total dust mass is M$_d = 1.6\times  10^7$ \msun. This high mass is inconsistent with current chemical evolution models (see next section). The least extreme value of cold dust temperatures providing a reasonable fit  (T$_c =$11 K, with T$_w$ = 19.6\,K) still yields a total dust mass M$_d = 2 \times 10^6$ \msun, which is six times higher than the value for single-temperature fit. 
Fitting the data with a  more sophisticated dust model to test alternative explanations is beyond the scope of this Letter; this analysis will be performed in a future work when the observations of the field are completed.
At this stage, given the low S/N of the detections at 500 \micron\ and the large error bars, it is difficult to discriminate between the different scenarios.

\section{Dust-to-gas mass ratios}

The dust-to-gas-mass ratio \D (M$_{\rm dust}$/M$_{\rm gas}$) not only provides information about the amount of metals that are locked in dust grains,
but also provides an indication of the star-formation history of a galaxy, reflecting the net balance between the formation and destruction  of dust \citep{2002A&A...388..439H}.
The ratio \D is known to correlate with the oxygen abundance \citep{1998ApJ...496..145L,2001MNRAS.328..223E,2002A&A...388..439H,2002MNRAS.335..753J}.
Here we check whether the gas and dust masses we obtained for these galaxies
in the Virgo cluster are compatible with the
predictions of dust formation models.

To derive $\mathcal D$,  we used neutral hydrogen masses available from the ALFALFA catalog
\citep[][see Table 1]{2007AJ....133.2569G,2008AJ....136..713K}.
The angular resolution of the 21-cm data (3$^{\prime\!}.3 \times 3^{\prime\!}.8$) includes all dust emission detected in these objects.  
We assume the total gas mass is given by the atomic component only:
independent of the lack of CO detections,
this assumption is justified by the low H$_2$-to-\hi\, mass fractions
expected in metal-poor environments \citep[e.g.,][]{2008ApJ...680.1083R,2009ApJ...693..216K,2009ApJ...699..850K}.

The resulting values of \D are reported in Table 1. 
Figure \ref{fig:d2g} shows \D for the Virgo BCDs
(filled circles) as a function of nebular oxygen abundance.
Also illustrated in Fig. \ref{fig:d2g} are data from the literature
\citep{2002MNRAS.335..753J,2005A&A...434..849H,2007ApJ...661..102W}.
Selected model predictions for the \D-metallicity correlation are also shown
\citep{2001MNRAS.328..223E,2002A&A...388..439H}.
The \citet{2002A&A...388..439H} models, shown as long-dashed and dot-dashed
curves, depend on the dust destruction
efficiency, which begins to be effective when the oxygen abundance $12+\log(O/H)$
 $\sim8$.
Were the interstellar dust mass an approximately constant
fraction of the ISM metal abundance,
as proposed by \citet{2001MNRAS.328..223E} and \citet{2002MNRAS.335..753J},
the relation between \D and $O/H$ would be linear.
A solid line indicates their prediction for dust production by SNe,
and the short-dashed line represents that by only evolved low- and intermediate-mass stars.
Alternatively, a non-linear trend is predicted by
\citet{1998ApJ...496..145L} and \citet{2002A&A...388..439H}
because of the additional effects of outflows and dust destruction efficiency.
The two most metal-poor objects plotted, I\,Zw\,18 and SBS\,0335$-$052 (at
$12+\log(O/H)\sim$7.2),
have a very low dust-to-gas ratio for their metallicity \citep{2005A&A...434..849H},
but are approximately consistent with linearity.

   \begin{figure}
\includegraphics[bb=10 0 415 200,width=8cm,clip]{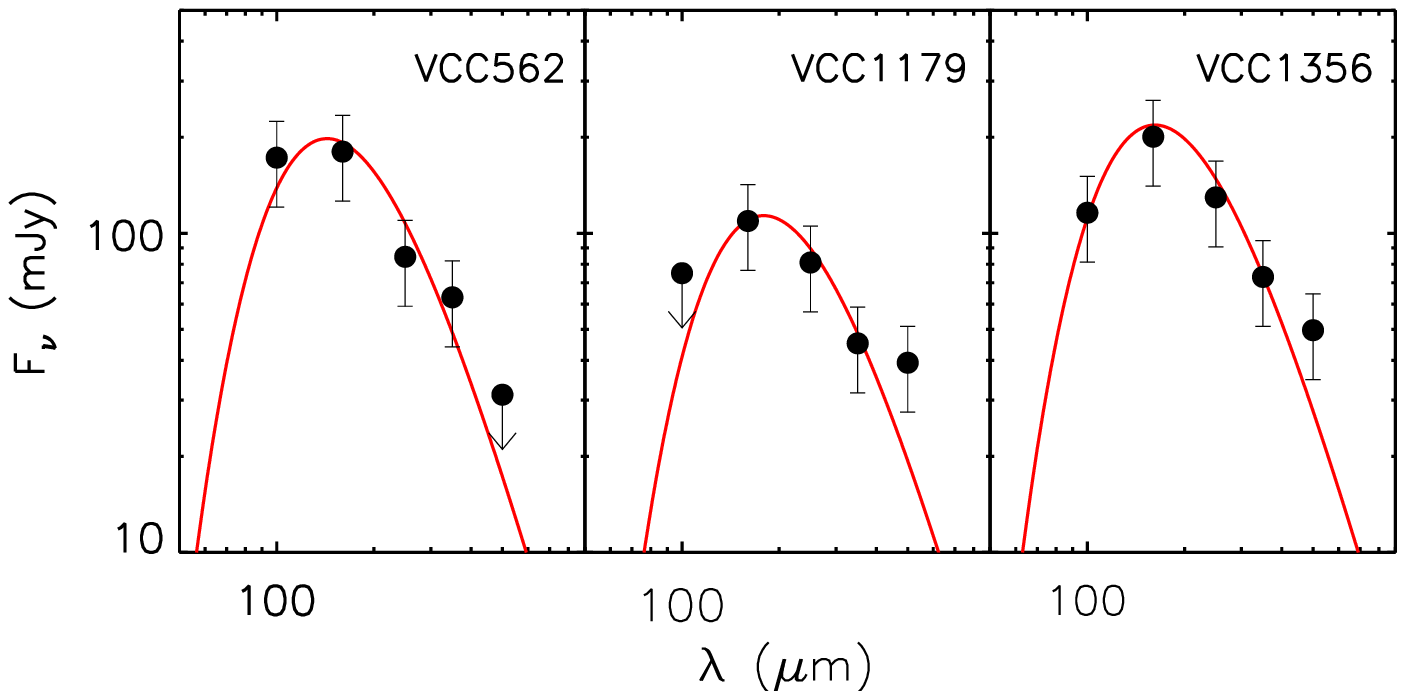}
\includegraphics[bb=23 8 410 208,width=8cm,clip]{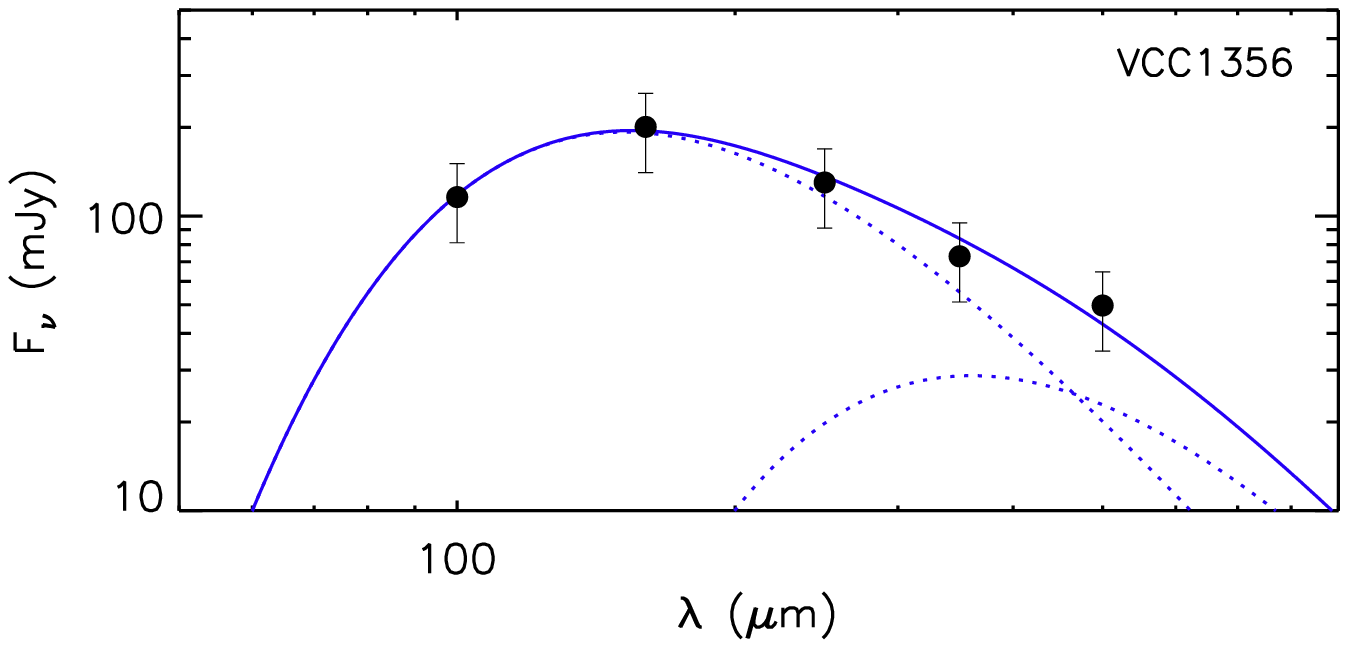}

   \caption{SEDs of the VCC dwarfs compared to a single modified black body with emissivity index $\beta = 2$ ({\em top}), and a two-component model fit with $\beta =2$, T$_w$= 19.1 K, T$_c= 8$ K for VCC\,1356 ({\em bottom}).
   }
              \label{fig:seds}%
    \end{figure}

   \begin{figure}
   \centering
   \includegraphics[bb=38 10 420 405,width=8cm,clip]{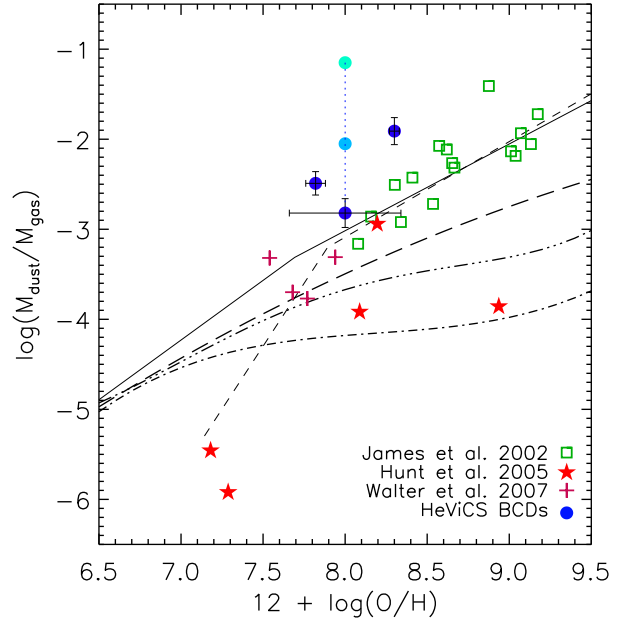}
   \caption{Dust-to-gas ratio \D versus oxygen abundance; 
HeViCS BCDs are shown as filled circles.
The vertical dotted line connects different dust-mass estimates for
VCC\,1356 as described in the text;
the top and middle point correspond to dust masses calculated for a two-component modified black body with T$_c$=8\,K and T$_c$=11\,K,
respectively, and the bottom point to a single-temperature fit.
Data from the literature are also plotted:
squares are from \citet{2002MNRAS.335..753J},
stars from \citet{2005A&A...434..849H}, and plus signs from \citet{2007ApJ...661..102W}.
The solid and short-dashed lines correspond to the models by
\citet{2002MNRAS.335..753J}, and the long-dashed and dot-dashed lines
to models 
by \citet{2002A&A...388..439H}. }
              \label{fig:d2g}%
    \end{figure}

Our estimates of \D for the three dwarfs, based on the single-temperature fit of the data,
are indeed consistent with what would be expected by the linear models. 
If these galaxies hosted a cold dust component with T$_c < 11$ K, as discussed in the previous section (and
shown in Fig. \ref{fig:d2g}), the corresponding \D would be much larger than the model predictions.
In this case, a significant missing gaseous component (H$_2$, cold H{\sc i}), a few times more massive than the amount of detected H{\sc i},
should be introduced to explain this discrepancy.


\section{Comparison with far-IR properties of known star-forming dwarfs}

The brightest and most metal-poor BCDs, such as IZw18 and SBS 0335-052E, have been studied with  {\em Spitzer},
and dust has been detected even in these low abundance environments \citep{2004ApJS..154..211H,2007ApJ...662..952W}.
While the SED peak of SBS 0335-052E is shifted to shorter wavelengths   \citep[$\lambda \sim$ 28 \micron; ][]{2004ApJS..154..211H}, I Zw 18 has also been detected at 70 \micron\ \citep{2007ApJ...662..952W,2008ApJ...678..804E}, 
suggesting that a cool dust component might be present in this galaxy.
Hunt et al. (2005) studied the global SEDs of seven BCDs showing that far-IR emission in these galaxies peaks at or shortward of $\sim$60 \micron.

However, the dwarfs detected with \hers\ are more ``quiescent'' than the typical BCDs analysed in these studies.
\citet{2007ApJ...661..102W} observed a sample of dwarf irregular galaxies in the M\,81 group with SFRs and metallicities more similar to the Virgo dwarfs. They derived dust masses and temperatures of the M\,81 dwarfs using only the 70 and 160 \micron\, {\em Spitzer}/MIPS bands.
Their dust-to-gas mass ratios, taking into account the total \hi\ mass of the galaxies, 
are displayed in Fig. \ref{fig:d2g} ({\em plus signs}), and
are about one order of magnitude smaller than those of the Virgo dwarfs with a similar metal abundance.
Finally, six Im-BCD galaxies in Virgo analysed with {\it ISO} data by \citet{2002ApJ...567..221P} show similar properties to the dwarfs in the current work. In particular, their 60-170 \micron\ SEDs indicate the presence of warm dust and a cooler component with a median temperature $\sim$18 K,
in agreement with our results. The resulting dust-to-gas mass ratios in a few cases are rather large,
$\ga$0.1. 


\section{Conclusions}

We have presented PACS and SPIRE observations of three star-forming dwarf galaxies in Virgo detected in the far-IR/submm regime
with the science demonstration phase data set for the HeViCS survey.
The data indicate the presence of cool dust with a temperature $\la$20 K and dust masses around 10$^5$ \msun.
We have discussed the possibility that these galaxies host an additional cold (T $\lesssim$ 10 K) component to explain the excess at 500 \micron\ in two of the dwarfs. However, the low S/N of the 500 \micron\ detections precludes us from drawing firm conclusions.
The completion of the full area of the survey will enable us to place more stringent constraints on the dust content of star-forming dwarf galaxies in a dense cluster environment.


\bibliographystyle{aa} 
\bibliography{14653bib} 

\end{document}